\documentclass[journal,twoside,web]{ieeecolor}
\usepackage{generic}
\usepackage{cite}
\usepackage{siunitx}
\usepackage{amsmath,amssymb,amsfonts}
\usepackage{algorithmic}
\usepackage{textcomp}

\usepackage{graphicx}
\usepackage{url}
\makeatletter
\let\NAT@parse\undefined
\makeatother
\usepackage{hyperref}
\hypersetup{
	colorlinks=true,
	linkcolor=blue,  
	urlcolor=blue,
	citecolor=blue
}
\def\BibTeX{{\rm B\kern-.05em{\sc i\kern-.025em b}\kern-.08em
    T\kern-.1667em\lower.7ex\hbox{E}\kern-.125emX}}
\usepackage{orcidlink}
\begin{document}
\title{Bounded Distribution Functions for Applied Physics, Especially Electron Device Simulation at Deep-Cryogenic Temperatures}
\author{\vspace{0.5cm}Arnout Beckers\,\orcidlink{0000-0003-3663-0824} \thanks{A. Beckers is with imec, Kapeldreef 75, 3001 Leuven, Belgium.}\thanks{(arnout.beckers@imec.be)}}

\maketitle

\begin{abstract}
Numerical underflow and overflow are major hurdles for rolling-out the modeling and simulation infrastructure for temperatures below about 50\,K. Extending the numeric precision is computationally intensive and thus best avoided. The root cause of these numerical challenges lies in the Fermi-Dirac, Bose-Einstein, and Boltzmann distribution functions. To tackle their extreme values, bounded distribution functions are proposed which are numerically safe in a given precision, yet identical to the standard distributions at the physical level. These functions can help to develop electron device models and TCAD software handling deep-cryogenic temperatures in the default double precision, to keep pace with the rapid experimental progress. More broadly, they can apply to other branches of applied physics with similar numerical challenges as well.
\end{abstract}


\section{Introduction}
\label{sec:introduction}
\IEEEPARstart{D}{ilution} refrigerators are becoming increasingly common nowadays in the pursuit of scalable quantum technologies \cite{chatterjee_semiconductor_2021}. Given the cost of this equipment and the long cool-down times, it is surprising that computer simulations are not used more often for device prototyping and shortening the time-to-application. The reason for this must be sought not only in the fact that some cryo-physics is missing in the simulation tools, but also, and most importantly for this article, in the numerical underflow and overflow in the default IEEE-754 double-precision arithmetic, which can cause immediate abortion of the program, or produce round-off errors prompting convergence issues in iterative solvers \cite{selberherr,kantner}. 

Consequently, extended precision is being explored in device models and TCAD \cite{richey,tedpaper,dhillon_tcad_2021}. However, it increases the run times by about $\times 10$ to $\times 100$ \cite{synopsys}, which is not acceptable for compact device models. Other strategies have also been applied, but with varying success rates: e.g., expanding logarithms, ramping the bias in a non-standard way, etc. \cite{jaeger_simulation_1980,akturk,mohiyaddin_multiphysics_2019,jin_considerations_2021,gao}. Highlighting some recent examples from the state-of-the-art : a dedicated charge-transport scheme allowed to simulate the MOSFET current characteristics down to \SI{4}{\kelvin} \cite{zlatan}; the terminal current in a diode could be simulated down to \SI{50}{\kelvin} using a two-step temperature-embedding strategy \cite{kantner}; a nanowire \emph{ab-initio} quantum transport simulation could be run at \SI{3}{\kelvin} \cite{wong}, and, an adaptive meshing strategy allowed to reach convergence in a gated quantum dot device down to \SI{1.4}{\kelvin} \cite{kriekouki} which was later extended down to \SI{100}{\milli\kelvin} \cite{beaudoin_robust_2022}. Yet, this is still one order of magnitude above the operating temperature of most semiconductor quantum devices ($\approx\SI{10}{\milli\kelvin}$). 

The aforementioned remedies do not remove the root cause of these numerical issues, which resides in the exponential temperature scaling of many semiconductor quantities, i.e., $\exp(-\Delta E/k_BT)$, which is the inevitable result of the underlying Fermi-Dirac (FD), Bose-Einstein (BE), and Boltzmann distribution functions ($\Delta E$ is an energy difference with respect to the Fermi level, $T$ is temperature, and $k_B$ is Boltzmann's constant). If $\Delta E=\SI{0.5}{\electronvolt}$ and $T=\SI{10}{\milli\kelvin}$ for example, it gives $\approx 10^{-250000}$, which transcends double ($10^{-308}$), quadruple ($\approx 10^{-4932}$), and octuple ($\approx 10^{-78913}$) precision.

This article proposes a bounded Boltzmann exponential, i.e., $\exp[S(\pm \Delta E / k_BT)]$, which can also be used in the FD and BE functions, where $S(\eta)$ follows from a semi-rigorous optimization problem using Lagrange multipliers including two additional constraints as compared to the standard derivation of the Boltzmann distribution. $S(\eta)$ avoids numerical issues in a given precision, while keeping $T$ in the simulations. Its use is similar to how variable-precision arithmetic (vpa) would be called in certain programming languages, i.e., $\exp\left[\mathrm{vpa}(\eta)\right]$, but, instead of converting $\eta$ into a higher precision format, it keeps the distribution functions in a chosen precision. 
\section{\label{sec:challenges}Numerical Challenges in the \\ Distribution Functions}
\begin{figure}[t]
	\centering
	\includegraphics[width=0.48\textwidth]{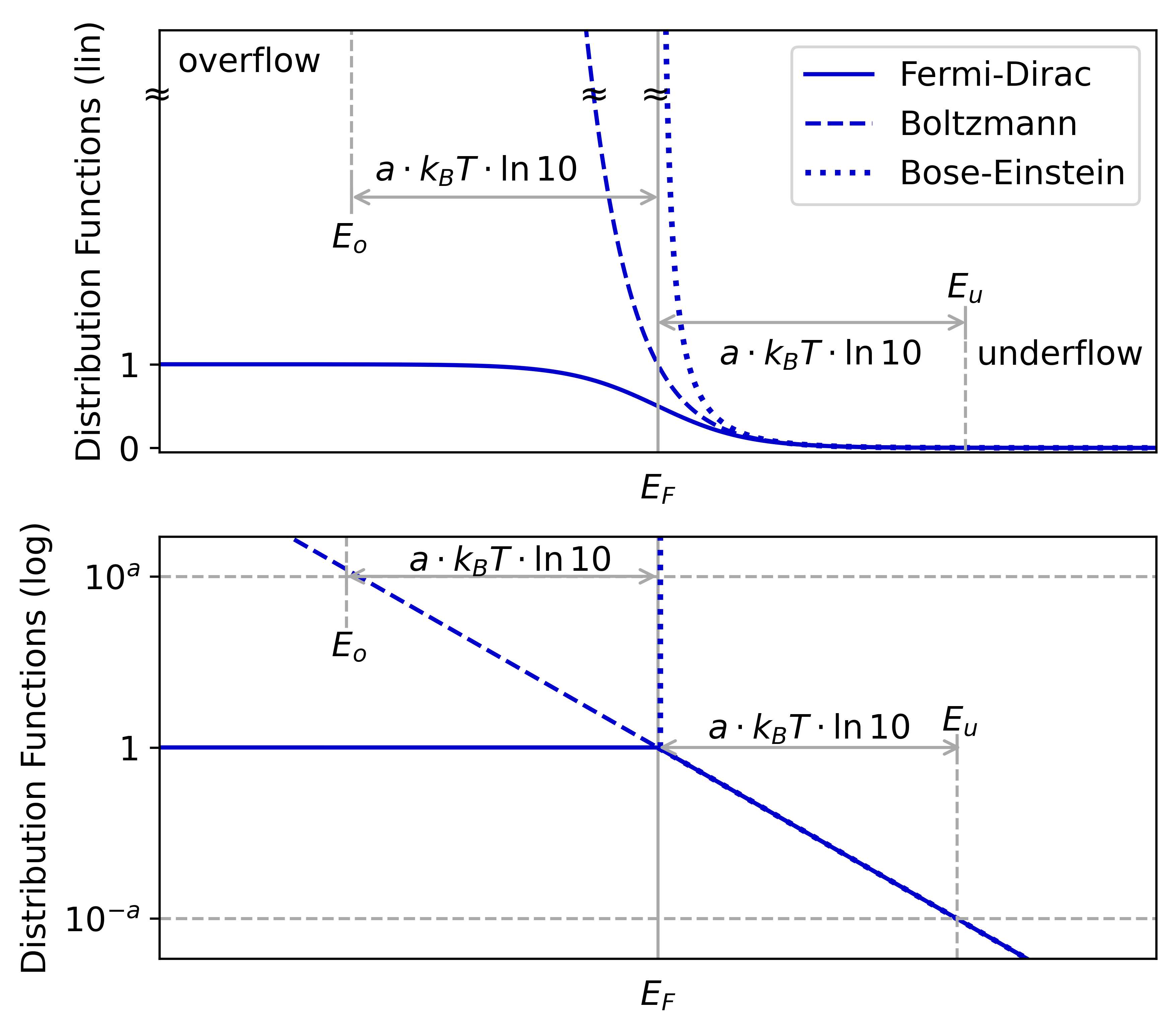}
	\vspace{-0.3cm}
	\caption{Boltzmann's distribution function underflows ($<10^{-a}$) above $E_u$ and overflows ($>10^{a}$) below $E_o$. The FD distribution underflows above $E_u$, but does not overflow. The BE distribution overflows above $E_F$ and underflows above $E_u$. The largest exponent \textquotedblleft $a$\textquotedblright \, is equal to 308 in double precision. At deep-cryogenic temperatures, $308 \cdot k_BT\cdot \ln10$ becomes smaller than typical bandgaps, causing numerical issues.}
	\label{fig:standard}
\end{figure}

Fig.\ref{fig:standard} shows the Boltzmann, FD, and BE functions in linear and logarithmic scales. The Boltzmann exponential,
\begin{equation}
	f_{B}=\exp\left(\frac{E_F-E}{k_BT}\right),
	\label{eq:b}
\end{equation}
underflows in a given precision if $f_B<10^{-a}$, where $a=\{308;\, 4932;\,78913\}$ for double, quadruple, and octuple precision, respectively. 
The underflow energy is $E_u=E_F+a\cdot k_BT\cdot\ln10$, and similarly, the overflow energy is $E_o=E_F-a\cdot k_BT\cdot\ln10$.  

\begin{figure}[t]
	\centering
	\includegraphics[width=0.48\textwidth]{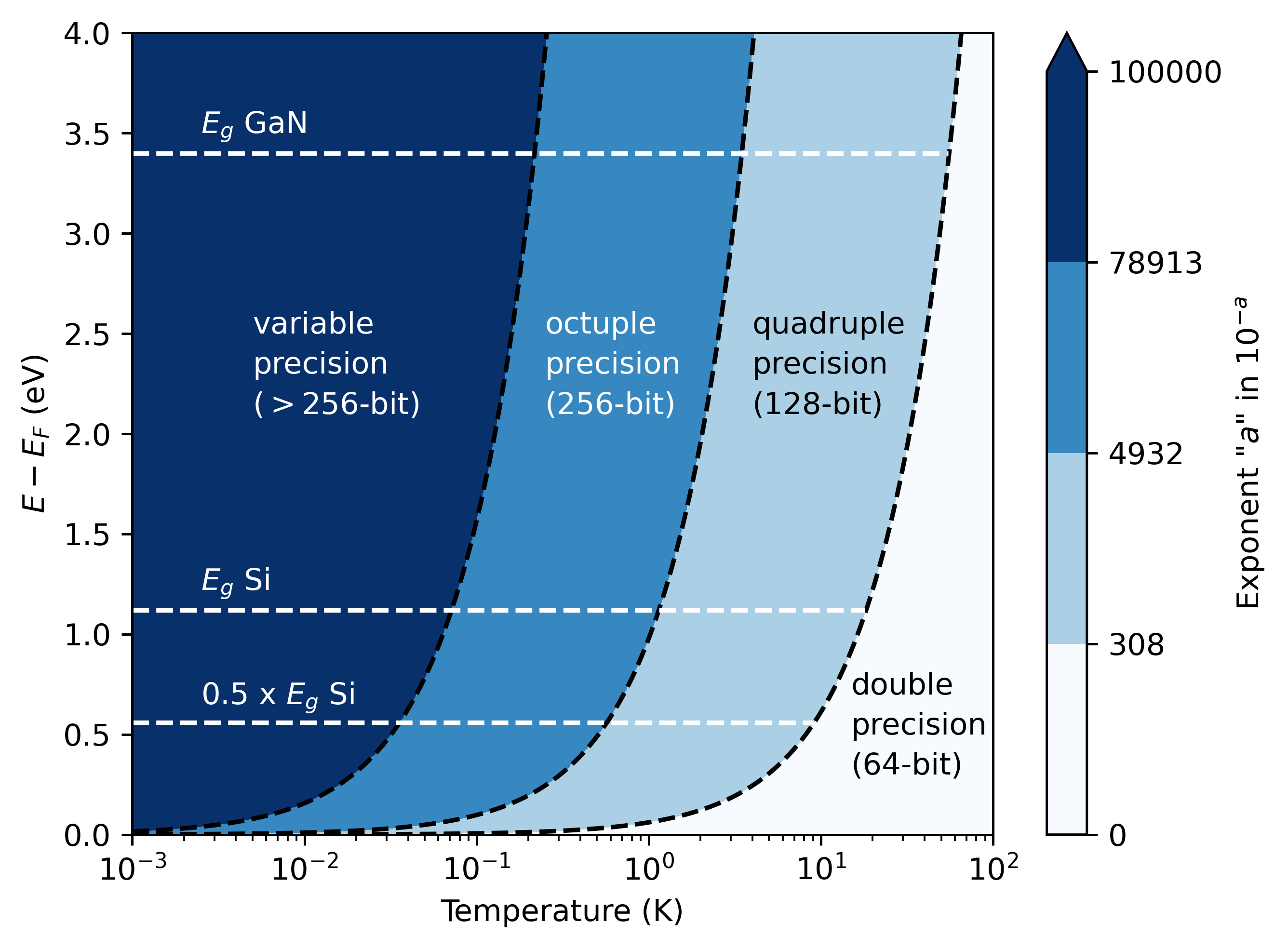}
	\vspace{-0.1cm}
	\caption{Map charting different zones depending on the numeric precision requirements imposed by (\ref{eq:b}) for a given $T$ and $\Delta E=E-E_F$. For silicon, double precision stops being sufficient around 10 to \SI{20}{\kelvin} if $E_F$ must be able to scan the bandgap without causing arithmetic underflow.}
	\label{fig:fd}
\end{figure}

Fig. \ref{fig:fd} charts the precision requirements obtained from $a=(E_u-E_F)/(k_BT\ln10)$. \SI{77}{\kelvin} still falls within the double-precision zone, shown in white, and therefore does not lead to numerical issues as long as the required energy range remains smaller than $\approx \SI{4}{\electronvolt}$, which is larger than typical semiconductor bandgaps (dashed white lines). For Si, the numerical issues will start around \SI{20}{\kelvin} if $E_F$ is required to scan the full bandgap, and around \SI{10}{\kelvin} for half the bandgap. For GaN, this will be around \SI{40}{\kelvin}. Underflow is known to be more severe for wide-bandgap semiconductors and at voltages for which the distance between $E_F$ and a band edge is large \cite{kantner,zlatan}. At \SI{10}{\milli\kelvin}, the energy window that can be simulated in double precision drops down to \SI{0.6}{\milli\electronvolt}. The rest of the bandgap will be numerically forbidden in double precision. Variable precision will be required, since even the octuple precision format is not sufficient to cover all possible positions of $E_F$ in the bandgap at such low temperatures.  

In non-equilibrium situations, the distribution functions might be broadened, increasing their effective temperature \cite{wong}. In other situations, the density-of-state functions might be smeared out increasing their effective overlap with the distribution function (e.g., band tails \cite{jap}). These are useful strategies but they are not generally applicable though. Furthermore, it has also been argued that the abrupt 0-K approximation of the FD function (i.e., Heaviside step function or metallic statistics) is suitable for deep-cryogenic device simulation \cite{aouad,catapano}. While this removes the danger of underflow because the exponential tail is removed, it also rejects the main temperature dependence, which is, nevertheless, an important physical variable for practical applications that wish to understand the differences in electrothermal behavior at temperatures between e.g., $\approx \SI{10}{\milli\kelvin}$ and \SI{10}{\kelvin}. Because these temperatures cannot all be categorized as \textquotedblleft \SI{0}{\kelvin}\textquotedblright, it is worthwhile to have available distribution functions which (i) keep $T$ in the simulations, (ii) are generally applicable, and (iii) live solely in the range covered by IEEE-754 double precision arithmetic. 
\newpage
\section{\label{sec:deriv}Numerically Safe Distribution Functions}
\subsection{Boltzmann Distribution Function}
The standard Boltzmann distribution distributes $N$ identical particles over the energy levels $\varepsilon_j$ such that the entropy is maximized for a fixed total energy ($E$) and fixed number of particles. $n_j$ is the number of particles in the energy level $\varepsilon_j$. This constrained optimization problem can be solved using Lagrange multipliers. Here, we add two constraints (besides particle and energy conservation) which assert a minimum and maximum occupation per energy level which are numerically safe in a chosen precision. It does not modify the physical solution of the problem for all practical purposes. 

Thus, we maximize the logarithm of the number of ways to place $n_j$ particle in $\varepsilon_j$,
\begin{equation}
	\ln(W)=N\ln(N)-\sum_jn_j\ln n_j
	\label{eq:lnw}
\end{equation}
subject to the following four constraints: 
\begin{enumerate}
	\item particles conservation $\Sigma_j n_j =N\Rightarrow \Sigma_j\delta n_j=0$,
	\item energy conservation $\Sigma_j \varepsilon_jn_j=E\Rightarrow \Sigma_j \varepsilon_j\delta n_j=0$,
	\item minimum occupation $p_j\geqslant 10^{-a}\Rightarrow \Sigma_j\delta n_j=0$,
	\item maximum occupation $p_j\leqslant 10^{a}\Rightarrow \Sigma_j\delta n_j=0$,
\end{enumerate}
where $p_j=n_j/N$ is the probability that $\varepsilon_j$ is occupied. The variation in $\ln(W)$ is then given by
\begin{equation}
	\delta \ln(W)=\sum_j\ln(n_j)\delta n_j=0,
\end{equation}
and, including the four constraints, this gives
$\sum_j\ln(n_j)\delta n_j+\alpha \sum_j\delta n_j+\beta \sum_j\varepsilon_j\delta n_j+(\gamma+\delta)\sum_j \delta n_j=0$, where $\alpha$, $\beta$, $\gamma$, and $\delta$ are Lagrange multipliers. This can be rewritten as 
\begin{equation}
	\Bigg\{\sum_j\ln(n_j)+\alpha+\beta \varepsilon_j+\gamma +\delta\Bigg\}\cdot \delta n_j=0.
\end{equation}

For arbitrariness of $\delta n_j$, we must have for all $j$ that 
\begin{equation}
	\ln(n_j)=-\alpha-\beta\varepsilon_j-\left(\gamma+\delta\right)
	\label{eq:nj}
\end{equation}
\begin{equation}
	\Leftrightarrow n_j=\exp\left(-\alpha-\beta \varepsilon_j-\gamma - \delta\right).
	\label{eq:mb}
\end{equation}
The first constraint is fulfilled if 
\begin{eqnarray}
	\sum_j \exp\left(-\alpha-\beta\varepsilon_j-\gamma-\delta\right)=N,
\end{eqnarray}
which gives 
\begin{equation}
	\ln\left(\frac{Z}{N}\right)=\alpha+\gamma+\delta,
	\label{eq:Z}
\end{equation}
where $Z=\sum_j\exp(-\beta\varepsilon_j)$ is the partition function.

Through (\ref{eq:mb}) and (\ref{eq:Z}), the probability $p_j=n_j/N$ can be expressed as 
\begin{equation}
	p_j=\frac{\exp(-\beta\varepsilon_j)}{Z}
	\label{eq:probability}
\end{equation}

Combining Boltzmann's entropy formula, $S=k_B\ln(W)$, and (\ref{eq:lnw}), gives 
\begin{equation}
	\sum_jn_j\ln(n_j)=\frac{-S+Nk_B\ln N}{k_B}
	\label{eq:njlnnj}
\end{equation}
With $\ln(n_j)$ from (\ref{eq:nj}), (\ref{eq:njlnnj}) turns into
\begin{equation}
	S=(\alpha+\gamma+\delta)k_BN+k_B\beta U+k_BN\ln(N)
\end{equation}
where $U=\sum_j\varepsilon_j$.

Hence the relation for $\beta$ is not modified by adding the two constraints, 
\begin{equation}
	\left(\frac{\partial S}{\partial U}\right)_V=\frac{1}{T}\Rightarrow \beta=\frac{1}{k_BT}.
\end{equation}
The extra constraints 3) and 4) are fulfilled if 
\begin{equation}
	10^{-a} \cdot N \leqslant \exp\left(-\alpha-\beta \varepsilon_j-\gamma - \delta\right) \leqslant 10^a \cdot N
	\label{eq:ineq}
\end{equation}
Using (\ref{eq:Z}) in (\ref{eq:ineq}) and working out, 
\begin{equation}
	-a\ln(10)-\ln(Z) \leqslant \beta \varepsilon_j \leqslant a\ln(10)-\ln(Z)
\label{eq:ineq2}
\end{equation}

\begin{figure}[t]
	\centering
	\includegraphics[width=0.49\textwidth]{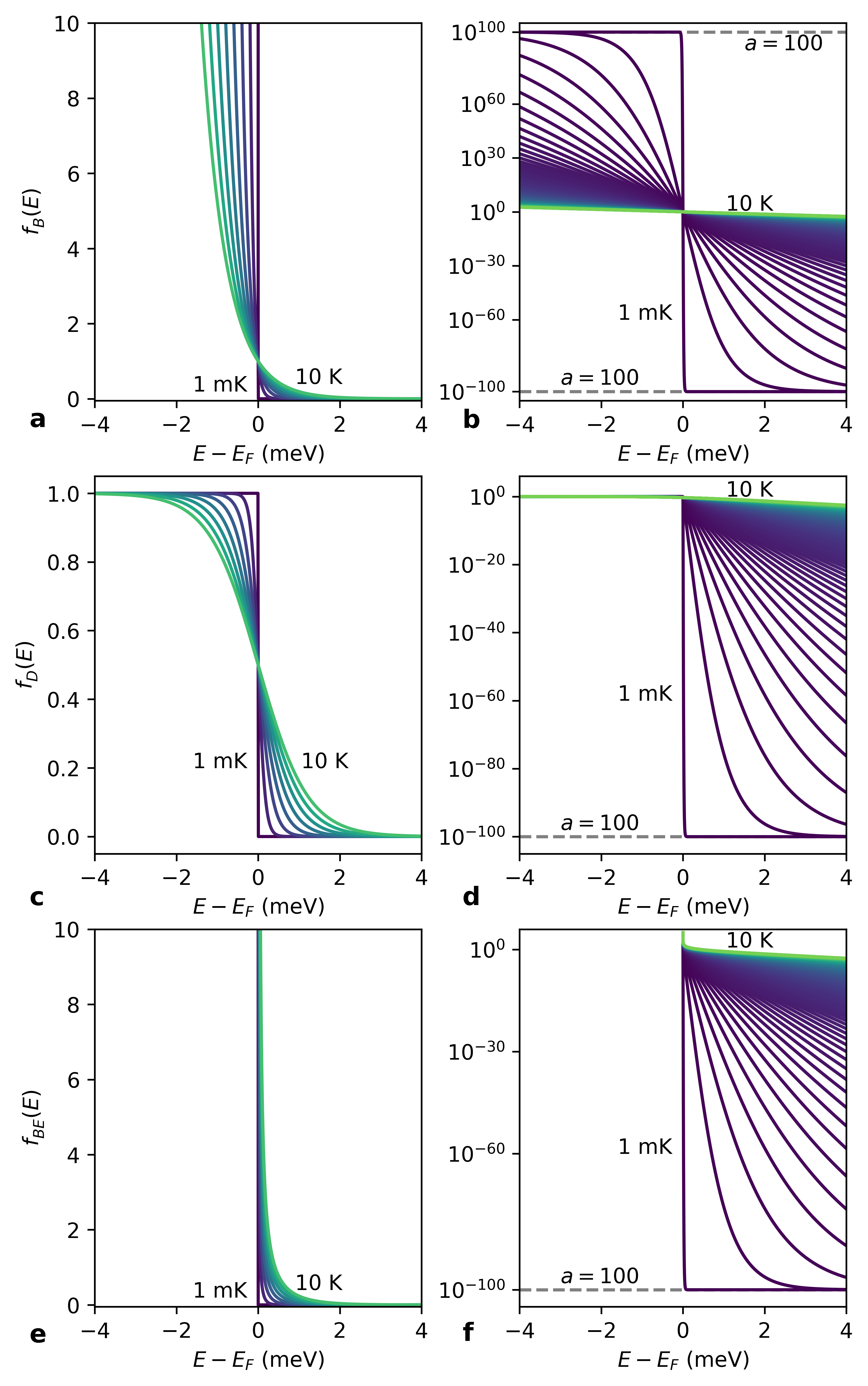}
	\vspace{-0.5cm}
	\caption{(a) Numerically safe Boltzmann distribution, (\ref{eq:nsb}) including (\ref{eq:Stanh}), in linear scale and (b) in log-scale. (c) Numerically safe FD function (\ref{eq:nsfd}) in linear scale, and (d) in log-scale. (e) Numerically safe BE function (\ref{eq:nsfd}) in linear scale, and (f) in log-scale. Expression (\ref{eq:S}) can be used instead of (\ref{eq:Stanh}) to obtain a faster transition to $10^{\pm 100}$ in log-scale.}
	\label{fig:ns}
\end{figure}

From the definition of the Fermi level $\varepsilon_F$, and (\ref{eq:probability}) it can further be obtained that 
\begin{equation}
p_j(\varepsilon_j=\varepsilon_F)=1 \Rightarrow \ln(Z)=-\beta\varepsilon_F
\label{eq:pj}
\end{equation}
Inserting (\ref{eq:pj}) in (\ref{eq:ineq2}) finally yields a bounded Boltzmann exponential 
\begin{eqnarray}
	\label{eq:cases}
	\begin{cases}
		\, p_j=\exp\left(-\eta\right),\\
		\, -a\ln(10) \leqslant \eta \leqslant a\ln(10),              
	\end{cases}
\end{eqnarray}
where $\eta=\beta(\varepsilon_j-\varepsilon_F )$. Thus the underflow and overflow energies that were mentioned in Section \ref{sec:challenges} are obtained rigorously from the optimization problem:
\begin{eqnarray}
	\begin{cases}
	\, \eta=a\ln(10)\Rightarrow \varepsilon_u = \varepsilon_F+\frac{1}{\beta}\cdot a\ln(10)\\
	\, \eta=-a\ln(10)\Rightarrow \varepsilon_o = \varepsilon_F-\frac{1}{\beta}\cdot a\ln(10)
	\end{cases}
\end{eqnarray}
\begin{figure}[t]
	\centering
	\includegraphics[width=0.5\textwidth]{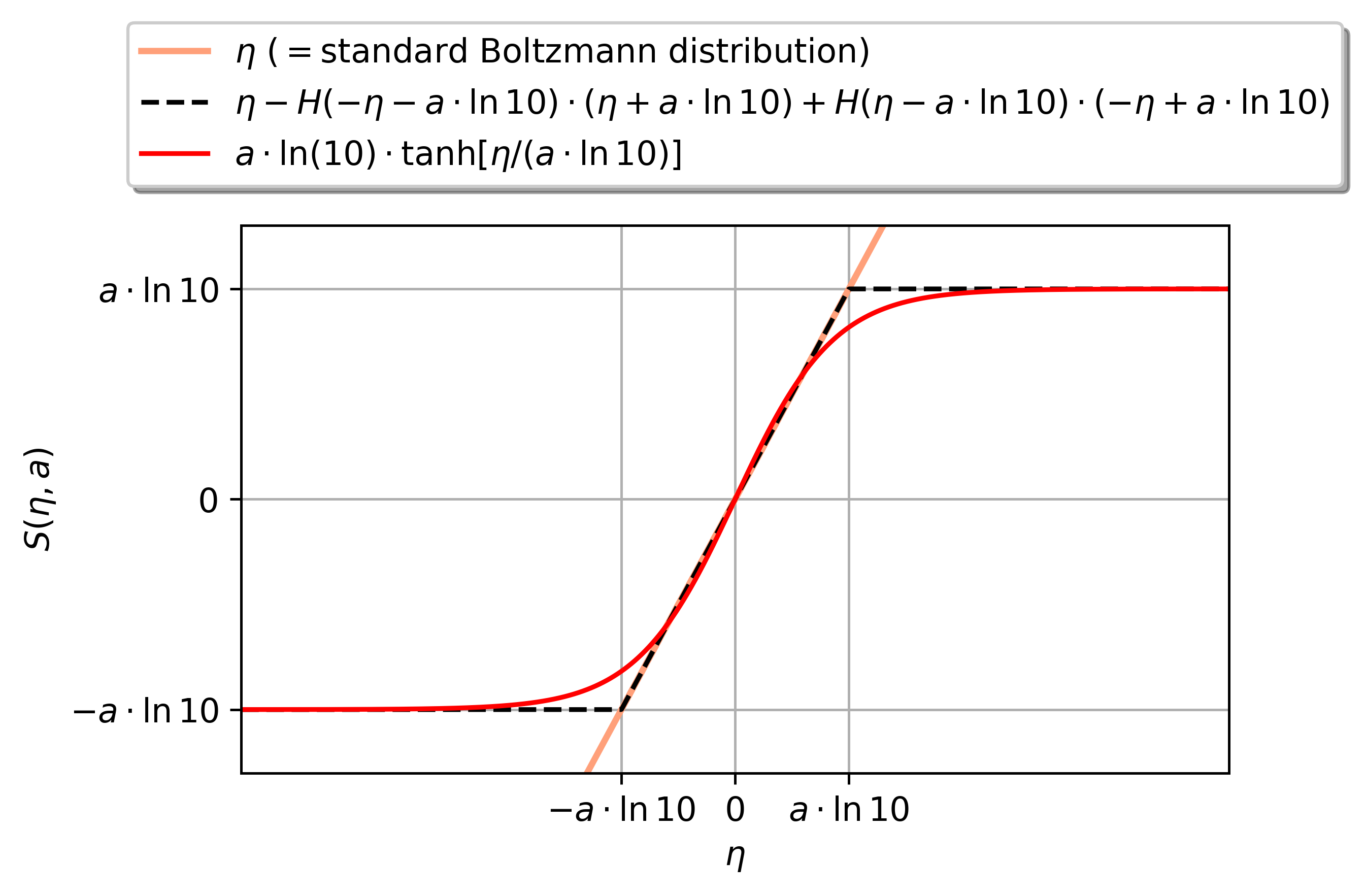}
	\vspace{-0.6cm}
	\caption{Function $S(\eta,a)$ in the Boltzmann exponent, see expressions (\ref{eq:S}) and (\ref{eq:Stanh}) in the main text.}
	\label{fig:S}
\end{figure}
Yet, the derivation does not tell us the shape of the function outside of its domain $\eta \in [-a\ln(10), a\ln(10)]$. The practical difficulty with (\ref{eq:cases}) is that it is not defined outside this domain. There are a couple of ways to analytically continue this function outside its domain in a semi-rigorous manner as long as the requirement $10^{-a}\leqslant p_j\leqslant 10^a$ remains fulfilled. For example, a continuous piecewise extension could be 
\begin{eqnarray}
\label{eq:piece}
\begin{cases}
	\, p_j=\exp(-\eta),  &  \eta \in [-a\ln10, a\ln10]\\
	\, 10^a,             &  \eta \in (-\infty, -a\ln10]\\
	\, 10^{-a},          &  \eta \in [a\ln10,+\infty)\\
\end{cases}
\end{eqnarray}
Equation (\ref{eq:piece}) is useful for numerical implementation, but for analysis it might be more convenient to have one expression that is valid over the whole domain from $-\infty$ to $+\infty$. Equation (\ref{eq:piece}) can be written more compactly as
\begin{equation}
	\label{eq:nsb}p_j(\eta,a)=\exp\left[-S\left(\eta,a\right)\right], \, \, \eta \in (-\infty, +\infty)
\end{equation}
where
\begin{eqnarray}
	S(\eta,a)=\eta-H(-\eta-a\cdot \ln10)\cdot \left(\eta+ a\cdot \ln10\right)\nonumber\\
	-H(\eta-a\cdot \ln10)\cdot \left(\eta- a\cdot \ln10\right),\label{eq:S}
\end{eqnarray}
which is plotted in Fig. \ref{fig:S}. If $\eta$ lies between $-a\cdot \ln(10)$ and $a\cdot \ln(10)$, then $S(\eta,a)=\eta$, which recovers the standard Boltzmann distribution. The disadvantage of (\ref{eq:S}) is the non-invertible nature of the function, which might be required during analytical model derivations in applied physics. Note, $S(\eta,a)$ has a piecewise sigmoid shape and can thus be approximated by an invertible hyperbolic tangent function, 
\begin{equation}
	S(\eta,a)\approx a\cdot\ln(10)\cdot \tanh\left(\frac{\eta}{a\cdot \ln10}\right),
	\label{eq:Stanh}
\end{equation}
tolerating some discrepancy around the corners and within the domain $\eta \in [-a\ln10, a\ln10]$ . Figs. \ref{fig:ns}(a) and \ref{fig:ns}(b) show that the safe Boltzmann exponential, (\ref{eq:nsb}) combined with (\ref{eq:S}) or (\ref{eq:Stanh}), does not underflow, nor overflow, in double precision if $a$ is set to a value below 308. The precision parameter \textquotedblleft $a$\textquotedblright \, can be chosen as an input by the user, e.g., $a=100$ will make the largest occupation number $10^{100}$ and the smallest $10^{-100}$.  
\subsection{Fermi-Dirac and Bose-Einstein Distribution Functions}
From (\ref{eq:nsb}), we infer numerically safe FD and BE functions,
\begin{equation}
	p_j(\eta,a)=\frac{1}{\exp\left[S\left(\eta,a\right)\right]\pm1}, 
	\label{eq:nsfd}
\end{equation}
where $+$ is for FD and $-$ for BE. Figs. \ref{fig:ns}(c)-(d) and \ref{fig:ns}(e)-(f) show the numerically safe FD and BE functions, respectively, for different temperatures down to \SI{1}{\milli\kelvin} in linear and logarithmic scales. The BE function is only defined above $E_o=E_F+k_BT\ln\left(1+10^{-a}\right)$. Note that these numerically safe distribution functions do not look different from the standard ones when plotted in linear scale. At the physical level, these bounded distribution functions are the same as the standard distribution functions, but the arithmetic underflow and overflow are avoided at the non-physical level.

It will be easiest to implement $S(\eta,a)$ in known analytical expressions, or use numerical integration, since $S(\eta,a)$ in the exponent can make it difficult to analytically integrate certain integrals. The intrinsic carrier concentration is the most commonly used example in the literature to illustrate the bad numerics at deep-cryogenic temperatures \cite{akturk,kantner,tedpaper,jin_considerations_2021}. However, it must be stressed that it is only one possible example among all the downstream semiconductor quantities impacted by the distribution functions. Other possible examples include transition rates in oxide-trap modeling \cite{michl}, dopant freezeout, Arrhenius-like temperature phenomena, FD or BE integrals, thermal properties, etc. Basically anywhere Boltzmann's exponential appears, the safe function $S(\eta,a)$ can be inserted in the exponent to stay within double precision (or any other precision chosen by the user). 

\subsection{Example Application : Intrinsic Carrier Concentration}
To avoid underflow in $n_i$, we write
\begin{equation}
	n_i^\prime=\sqrt{N_cN_v}\cdot\exp\left[S\left(\frac{-E_g}{2k_BT},a\right)\right],
	\label{eq:nsni}
\end{equation}
where $S(\eta,a)$ is given either by (\ref{eq:S}) or (\ref{eq:Stanh}), and $N_c$ and $N_v$ are the regular effective density-of-states, which scale as $\propto T^{3/2}$ and thus do not constitute a numerical challenge. 
\begin{figure}[t]
	\centering
	\includegraphics[width=0.48\textwidth]{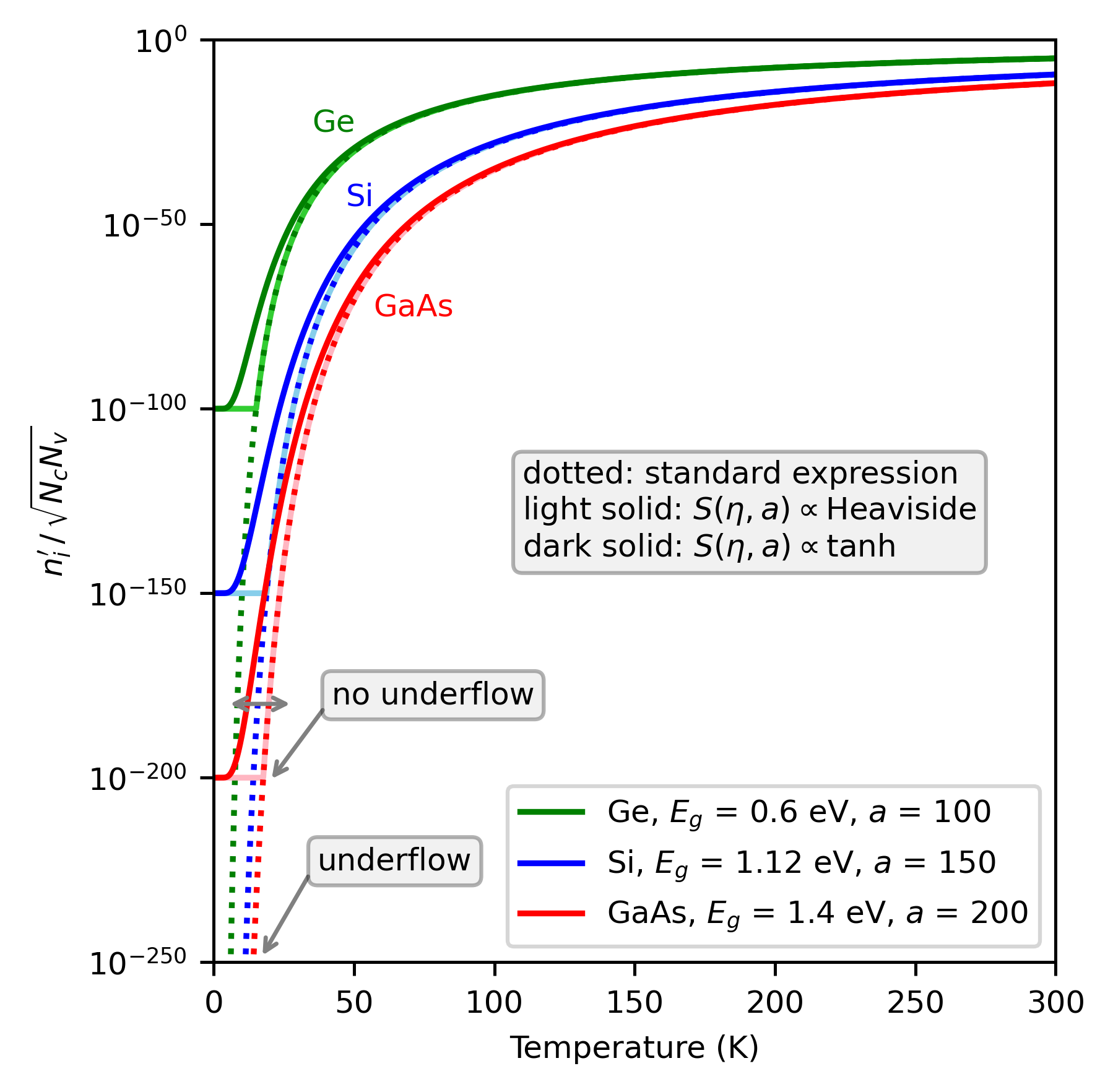}
	\caption{Intrinsic carrier concentration (\ref{eq:nsni}) does not underflow in double precision if $a<308$. Using $S(\eta,a)$ with the Heaviside functions (\ref{eq:S}) is more accurate at low density. The hyperbolic tangent approximation (\ref{eq:Stanh}) has a slower roll-off and deviates from the standard expression (annotated with a horizontal double arrow for GaAs).} 
\label{fig:ni}
\end{figure}

This \textquotedblleft numerically safe $n_i$\textquotedblright \, is plotted in Fig. \ref{fig:ni} for different $a$, demonstrating that $n_i^\prime$ has ceased to underflow in double precision for Si, Ge, and GaAs ($a$ freely chosen for each). Light solid lines use the Heaviside $S$ (\ref{eq:S}), which gives more accurate results at low density than the approximative hyperbolic tangent (\ref{eq:Stanh}), which has a slower roll-off. In practice, however, this discrepancy (indicated with the horizontal double arrow) might be tolerable for the sake of having an invertible function $S$, since the absolute values are extremely low. 

\section{A Note on the Steepness of the\\ Distribution Functions}
Besides underflow and overflow, the remaining issue introduced by (\ref{eq:b}) is the steepness of the exponential tail, imposing an extreme sensitivity to variations in $E_F$. The sensitivity scales as $1/T$, i.e., 
\begin{equation}
	\left|\frac{\partial f_B(E,E_F)}{\partial E}\right|_{E_F}=\frac{1}{k_BT},
	\label{eq:sensitivity}
\end{equation}
which can lead to sharp internal layers in devices requiring ever finer meshing at lower temperatures. The minimum mesh size is set by the Debye length $L_D=\sqrt{\varepsilon k_BT/(q^2N)}$ \cite{vasileska}. This strong sensitivity can result in non-convergence in Poisson's equation using an iterative solver such as Newton-Raphson. The proposed Boltzmann tail has stopped to underflow (and overflow) but still has an extremely steep slope. This slope is an inherent feature of $T$, so it was retained in (\ref{eq:nsb}) and (\ref{eq:nsfd}). As such, these bounded distribution functions still need to be paired with a dedicated meshing strategy \cite{beaudoin_robust_2022} and/or a suitable modification of the solution variable \cite{zlatan} in order to reach convergence at deep-cryogenic/sub-Kelvin temperatures, at high biases, and/or in wide-bandgap materials. In a companion paper, the combined strategy of using the bounded distribution function with a transformation of the solution variable in Poisson's equation, has already been successfully applied to achieve convergence in the electrostatic simulation of a $p$-$n$ diode down to \SI{1}{\micro\kelvin} \cite{beckers_diode}. Albeit still a basic simulation, this counts as the first-ever reported semiconductor device simulation at such low temperature. 

\section{\label{sec:conclusions}Conclusions}
\begin{itemize}
	\item Low-temperature device simulations are hampered by the emergence of numerical issues originating from the temperature scaling of the Boltzmann exponential, which also arises in the tails of the Fermi-Dirac and Bose-Einstein functions, $f(E)$, and therefore in all semiconductor quantities based on the form $\int g(E)\cdot f(E)\cdot dE$, where $g(E)$ is typically a density-of-states function. The often quoted $n_i$ is only one example. 
	\item Extending the numeric precision beyond the standard 64-bit is slow and not always available. More than octuple precision (256-bit) would be required to simulate devices at the often-used experimental temperature of $\approx \SI{10}{\milli\kelvin}$. We charted the limits of each precision format over temperature and energy.
	\item Two additional constraints in the derivation of the standard Boltzmann distribution (besides particle and energy conservation), led to an upper- and lower-bounded Boltzmann exponential, which was also implemented in the Fermi-Dirac and Bose-Einstein distributions.
	\item These dedicated distribution functions can help to extend TCAD simulators and physics-based device models into the deep-cryogenic temperature regime, while maintaining the standardized and fast double precision. 
\end{itemize}  

\bibliographystyle{ieeetran}
\bibliography{bounded}

\begin{thebibliography}{10}
\providecommand{\url}[1]{#1}
\csname url@samestyle\endcsname
\providecommand{\newblock}{\relax}
\providecommand{\bibinfo}[2]{#2}
\providecommand{\BIBentrySTDinterwordspacing}{\spaceskip=0pt\relax}
\providecommand{\BIBentryALTinterwordstretchfactor}{4}
\providecommand{\BIBentryALTinterwordspacing}{\spaceskip=\fontdimen2\font plus
\BIBentryALTinterwordstretchfactor\fontdimen3\font minus
  \fontdimen4\font\relax}
\providecommand{\BIBforeignlanguage}[2]{{%
\expandafter\ifx\csname l@#1\endcsname\relax
\typeout{** WARNING: IEEEtran.bst: No hyphenation pattern has been}%
\typeout{** loaded for the language `#1'. Using the pattern for}%
\typeout{** the default language instead.}%
\else
\language=\csname l@#1\endcsname
\fi
#2}}
\providecommand{\BIBdecl}{\relax}
\BIBdecl

\bibitem{chatterjee_semiconductor_2021}
A.~Chatterjee, P.~Stevenson, S.~De~Franceschi, A.~Morello, N.~P. de~Leon, and
  F.~Kuemmeth, ``Semiconductor qubits in practice,'' \emph{Nature Reviews
  Physics}, vol.~3, no.~3, pp. 157--177, Mar. 2021,
  \href{http://www.nature.com/articles/s42254-021-00283-9}{doi:10.1038/s42254-021-00283-9}.

\bibitem{selberherr}
S.~Selberherr, ``{MOS} device modeling at 77 {K},'' \emph{IEEE Transactions on
  Electron Devices}, vol.~36, no.~8, pp. 1464--1474, Aug. 1989,
  \href{https://ieeexplore.ieee.org/document/30960}{doi:10.1109/16.30960}.

\bibitem{kantner}
M.~Kantner and T.~Koprucki, ``Numerical simulation of carrier transport in
  semiconductor devices at cryogenic temperatures,'' \emph{Optical and Quantum
  Electronics}, vol.~48, no.~12, Dec. 2016,
  \href{http://link.springer.com/10.1007/s11082-016-0817-2}{doi:10.1007/s11082-016-0817-2}.

\bibitem{richey}
\BIBentryALTinterwordspacing
D.~M. Richey, J.~D. Cressler, and R.~C. Jaeger, ``Numerical simulation of
  {SiGe} {HBT}'s at cryogenic temperatures,'' \emph{Le Journal de Physique IV},
  vol.~04, pp. C6--127--C6--132, Jun. 1994. [Online]. Available:
  \url{http://www.edpsciences.org/10.1051/jp4:1994620}
\BIBentrySTDinterwordspacing

\bibitem{tedpaper}
A.~Beckers, F.~Jazaeri, and C.~Enz, ``Cryogenic {MOS} {Transistor} {Model},''
  \emph{IEEE Transactions on Electron Devices}, vol.~65, no.~9, pp. 3617--3625,
  Sep. 2018,
  \href{https://ieeexplore.ieee.org/document/8424046/}{doi:10.1109/TED.2018.2854701}.

\bibitem{dhillon_tcad_2021}
P.~Dhillon, N.~C. Dao, P.~H.~W. Leong, and H.~Y. Wong, ``{TCAD} {Modeling} of
  {Cryogenic} {nMOSFET} {ON}-{State} {Current} and {Subthreshold} {Slope},'' in
  \emph{2021 {International} {Conference} on {Simulation} of {Semiconductor}
  {Processes} and {Devices} ({SISPAD})}.\hskip 1em plus 0.5em minus 0.4em\relax
  Dallas, TX, USA: IEEE, Sep. 2021, pp. 255--258,
  \href{https://ieeexplore.ieee.org/document/9592586/}{doi:10.1109/SISPAD54002.2021.9592586}.

\bibitem{synopsys}
Synopsys, ``Sentaurus™ {Device} {User} {Guide},'' p. 213.

\bibitem{jaeger_simulation_1980}
R.~Jaeger and F.~Gaensslen, ``Simulation of impurity freezeout through
  numerical solution of {Poisson}'s equation with application to {MOS} device
  behavior,'' \emph{IEEE Transactions on Electron Devices}, vol.~27, no.~5, pp.
  914--920, May 1980, doi:
  \href{http://ieeexplore.ieee.org/document/1480749/}{10.1109/T-ED.1980.19956}.

\bibitem{akturk}
A.~Akturk, M.~Holloway, S.~Potbhare, D.~Gundlach, B.~Li, N.~Goldsman,
  M.~Peckerar, and K.~P. Cheung, ``Compact and {Distributed} {Modeling} of
  {Cryogenic} {Bulk} {MOSFET} {Operation},'' \emph{IEEE Transactions on
  Electron Devices}, vol.~57, no.~6, pp. 1334--1342, Jun. 2010, doi:
  \href{http://ieeexplore.ieee.org/document/5456141/}{10.1109/TED.2010.2046458}.

\bibitem{mohiyaddin_multiphysics_2019}
F.~A. Mohiyaddin, B.~Chan, T.~Ivanov, A.~Spessot, P.~Matagne, J.~Lee,
  B.~Govoreanu, I.~P. Radu, G.~Simion, N.~I.~D. Stuyck, R.~Li, F.~Ciubotaru,
  G.~Eneman, F.~M. Bufler, S.~Kubicek, and J.~Jussot, ``Multiphysics
  {Simulation} \& {Design} of {Silicon} {Quantum} {Dot} {Qubit}
  {Devices}.''\hskip 1em plus 0.5em minus 0.4em\relax IEEE, Dec. 2019, pp.
  39.5.1--39.5.4,
  \href{https://ieeexplore.ieee.org/document/8993541/}{10.1109/IEDM19573.2019.8993541}.

\bibitem{jin_considerations_2021}
S.~Jin, A.-T. Pham, W.~Choi, M.~A. Pourghaderi, U.~Kwon, and D.~S. Kim,
  ``Considerations for {DD} {Simulation} at {Cryogenic} {Temperature},'' in
  \emph{2021 {International} {Conference} on {Simulation} of {Semiconductor}
  {Processes} and {Devices} ({SISPAD})}.\hskip 1em plus 0.5em minus 0.4em\relax
  Dallas, TX, USA: IEEE, Sep. 2021, pp. 251--254,
  \href{https://ieeexplore.ieee.org/document/9592572/}{doi:10.1109/SISPAD54002.2021.9592572}.

\bibitem{gao}
X.~Gao, E.~Nielsen, R.~P. Muller, R.~W. Young, A.~G. Salinger, N.~C. Bishop,
  M.~P. Lilly, and M.~S. Carroll, ``Quantum computer aided design simulation
  and optimization of semiconductor quantum dots,'' \emph{Journal of Applied
  Physics}, vol. 114, no.~16, p. 164302, Oct. 2013, doi:
  \href{http://aip.scitation.org/doi/10.1063/1.4825209}{10.1063/1.4825209}.

\bibitem{zlatan}
\BIBentryALTinterwordspacing
Z.~Stanojevic, J.~M. Gonzalez~Medina, F.~Schanovsky, and M.~Karner,
  ``{Quasi-Fermi-Based Charge Transport Scheme for Device Simulation in
  Cryogenic, Wide-Band-Gap, and High-Voltage Applications},'' \emph{{Preprint
  Submitted to Transactions on Electron Devices}}. [Online]. Available:
  \url{https://doi.org/10.36227/techrxiv.21132637.v1}
\BIBentrySTDinterwordspacing

\bibitem{wong}
\BIBentryALTinterwordspacing
T.~Jiao and H.~Y. Wong, ``Robust cryogenic ab-initio quantum transport
  simulation for {$L_G$} = 10 nm nanowire,'' \emph{Solid-State Electronics},
  vol. 197, p. 108440, Nov. 2022. [Online]. Available:
  \url{https://linkinghub.elsevier.com/retrieve/pii/S003811012200212X}
\BIBentrySTDinterwordspacing

\bibitem{kriekouki}
\BIBentryALTinterwordspacing
I.~Kriekouki, F.~Beaudoin, P.~Philippopoulos, C.~Zhou, J.~Camirand~Lemyre,
  S.~Rochette, S.~Mir, M.~J. Barragan, M.~Pioro-Ladrière, and P.~Galy,
  ``\BIBforeignlanguage{en}{Interpretation of 28 nm {FD}-{SOI} quantum dot
  transport data taken at 1.4 {K} using {3D} quantum {TCAD} simulations},''
  \emph{\BIBforeignlanguage{en}{Solid-State Electronics}}, vol. 194, p. 108355,
  Aug. 2022. [Online]. Available:
  \url{https://linkinghub.elsevier.com/retrieve/pii/S0038110122001277}
\BIBentrySTDinterwordspacing

\bibitem{beaudoin_robust_2022}
\BIBentryALTinterwordspacing
F.~Beaudoin, P.~Philippopoulos, C.~Zhou, I.~Kriekouki, M.~Pioro-Ladrière,
  H.~Guo, and P.~Galy, ``Robust technology computer-aided design of gated
  quantum dots at cryogenic temperature,'' \emph{Applied Physics Letters}, vol.
  120, no.~26, p. 264001, Jun. 2022. [Online]. Available:
  \url{https://aip.scitation.org/doi/10.1063/5.0097202}
\BIBentrySTDinterwordspacing

\bibitem{jap}
\BIBentryALTinterwordspacing
A.~Beckers, D.~Beckers, F.~Jazaeri, B.~Parvais, and C.~Enz, ``Generalized
  {Boltzmann} relations in semiconductors including band tails,'' \emph{Journal
  of Applied Physics}, vol. 129, no.~4, p. 045701, Jan. 2021. [Online].
  Available: \url{http://aip.scitation.org/doi/10.1063/5.0037432}
\BIBentrySTDinterwordspacing

\bibitem{aouad}
M.~Aouad, T.~Poiroux, S.~Martinie, F.~Triozon, M.~Vinet, and G.~Ghibaudo,
  ``Poisson-{Schrödinger} simulation and analytical modeling of inversion
  charge in {FDSOI} {MOSFET} down to 0 {K} – {Towards} compact modeling for
  cryo {CMOS} application,'' \emph{Solid-State Electronics}, vol. 186, p.
  108126, Dec. 2021,
  \href{https://linkinghub.elsevier.com/retrieve/pii/S0038110121001696}{doi:10.1016/j.sse.2021.108126}.

\bibitem{catapano}
E.~Catapano, M.~Cassé, F.~Gaillard, S.~{de Franceschi}, T.~Meunier, M.~Vinet,
  and G.~Ghibaudo, ``{TCAD} {Simulations} of {FDSOI} devices down to {Deep}
  {Cryogenic} {Temperature},'' \emph{Solid-State Electronics}, p. 108319, 2022,
  doi:
  \href{https://www.sciencedirect.com/science/article/pii/S0038110122000910}{10.1016/j.sse.2022.108319}.

\bibitem{michl}
\BIBentryALTinterwordspacing
J.~Michl, A.~Grill, D.~Claes, G.~Rzepa, B.~Kaczer, D.~Linten, I.~Radu,
  T.~Grasser, and M.~Waltl, ``Quantum {Mechanical} {Charge} {Trap} {Modeling}
  to {Explain} {BTI} at {Cryogenic} {Temperatures}.''\hskip 1em plus 0.5em
  minus 0.4em\relax IEEE, Apr. 2020, pp. 1--6. [Online]. Available:
  \url{https://ieeexplore.ieee.org/document/9128349/}
\BIBentrySTDinterwordspacing

\bibitem{vasileska}
\BIBentryALTinterwordspacing
D.~Vasileska, S.~M.~Goodnick, and G.~Klimeck, \emph{Computational
  {Electronics}: {Semiclassical} and {Quantum} {Device} {Modeling} and
  {Simulation}}, 1st~ed.\hskip 1em plus 0.5em minus 0.4em\relax CRC Press, Dec.
  2017. [Online]. Available:
  \url{https://www.taylorfrancis.com/books/9781420064841}
\BIBentrySTDinterwordspacing

\bibitem{beckers_diode}
A.~Beckers, ``\BIBforeignlanguage{en}{{Robust Simulation of Poisson's Equation
  in a P-N Diode Down to 1\,$\mu$K}},'' Dec. 2022.

\end{thebibliography}

\end{document}